\documentclass[oldversion,letter]{aa}

\usepackage{graphicx}
\usepackage{txfonts}
\usepackage{natbib}
\usepackage{color}
\usepackage{multirow} 	
\usepackage{relsize}	
\usepackage{balance}

\def\sss{\scriptscriptstyle}
\def\U{{\sss \!U}}
\def\L{{\sss \!L}}
\def\K{{\sss \!K}}

\def\S{{\sss \!S}}
\def\I{{\sss \!I}}
\def\C{{\sss \!C}}

\def\O{{\sss \!O}}
\def\R{{\sss \!R}}

\def\nur{\nu_\mathrm{r}}
\def\nuv{\nu_\theta}
\def\nuL{\nu_\L}
\def\nuU{\nu_\U}
\def\nuK{\nu_\K}

\def\nuISCO{\nu_{\I\S\C\O}}
\def\ISCO{\I\S\C\O}
\def\RISCO{\R\I\S\C\O}
\def\rISCO{r_{\ISCO}}
\def\rRISCO{r_{\RISCO}}
\def\nuRISCO{\nu_{\R\I\S\C\O}}

\def\xmmu{XMMUJ134736.6+173403}

\begin{document}

\title{Mass of the active galactic nucleus black hole  \xmmu}

\author
{K. Goluchov\'a\inst{1}, G. T\"or\"ok\inst{1}, E. \v{S}r\'amkov\'a\inst{1}, Marek A. Abramowicz\inst{1}, Z. Stuchl\'{\i}k\inst{1},  Ji\v{r}\'{\i} Hor\'{a}k\inst{2}}

\institute{
$^1$ Institute of Physics, Research Centre for Computational Physics and Data Processing, Research Centre for Theoretical Physics and Astrophysics,
Faculty of Philosophy \& Science, Silesian University in Opava,  Bezru\v{c}ovo n\'am.~13, CZ-746\,01 Opava, Czech Republic\\
$^2$ Astronomical Institute, Bo\v{c}n\'{\i} II 1401/2a, CZ-14131 Praha 4-Spo\v{r}ilov, Czech Republic
}


\authorrunning{K. Goluchov\'a et al.}
\titlerunning{QPOs and mass of AGN BHs}
 
\date{Received 4/12/2018,    Accepted 09/01/2019}

\abstract
{A recent study of the X-ray source \xmmu~ has revealed a strong quasi-periodic modulation of the X-ray flux. The observation of two quasiperiodic oscillations (QPOs) that occur on a daily timescale and exhibit a 3:1 frequency ratio  strongly supports the evidence for the presence of an active galactic nucleus black hole (AGN BH). Assuming an orbital origin of QPOs, we calculated the upper and lower limit on AGN BH mass $M$ and found $M\approx 10^7-10^9M_{\sun}$. When we compare this to mass estimates of other sources, \xmmu~ appears to be the most massive source with commensurable QPO frequencies, and its mass represents the current observational upper limit on AGN BH mass based on QPO observations. We note that it will be crucial for the falsification of particular resonance models of QPOs whether only a single QPO with a frequency that completes the harmonic sequence $3\!:\!2\!:\!1$ is found in this source, or if a new different pair of QPOs with frequencies in the $3:\!2$ ratio is found. The former case would agree with the prediction of the $3\!:\!2$ epicyclic resonance model and BH mass $M\approx(5a^2 + 8a + 8)\times 10^{7}M_{\sun}$, where $a$ is a dimensionless BH spin.
}
\maketitle

\section{Introduction}
\label{section:introduction}

The X--ray {power density spectra (PDS)} of several Galactic black hole (BH) systems displays so-called high-frequency quasi-periodic oscillations (HF QPOs) within the range of $40-450$Hz. It has been noted that their frequencies lie within the range corresponding to timescales of orbital motion in the vicinity of a BH. This coincidence is thought to represent a strong indication that the observed signal originates in the innermost parts of an accretion disk. This hypothesis also finds support in the field of Fourier-resolved spectroscopy \citep[e.g.,][]{gilf-etal:2000,mcc-rem:2006,kli:2006}.


{Detections of elusive HF QPO peaks in Galactic microquasars are often reported at rather constant frequencies that are characteristic for a given source. It has been found that they usually appear in ratios of small natural numbers \citep{abr-klu:2001,rem-etal:2002, mcc-rem:2006}, typically in a {$3\!:\!2$} ratio \citep[see, however,][]{bel-etal:2012,bel-alt:2013,var-rod:2018}. The evidence for the rational frequency ratios was also discussed in the context of neutron star (NS) QPOs. In the NS sources, clustering of twin-peak QPO detections most frequently arises as a result of the weakness of (one or both) QPOs outside the limited range of QPO frequencies \citep[frequency ratio; see][]{abr-etal:2003a, bel-etal:2005, bel-etal:2007, tor-etal:2008b, tor-etal:2008a, bar-bou:2008, bou-etal:2010}.}

The {$3\!:\!2$} frequencies observed in the Galactic microquasars are well matched by the relation \citep[][]{mcc-rem:2006}
\begin{equation}
\label{equation:bestfit}
\nuU = \frac{2.8\mathrm{kHz}}{M^{*}},
\end{equation}
where $\nuU$ is the higher of the two frequencies that form the $3\!:\!2$ ratio, $R=\nuU/\nuL= 3/2$, {and the BH gravitational mass is given as} $M^{*}=M/M_{\odot}$.


{\cite{abr-etal:2004:ApJ:} suggested that detections of ${3\!:\!2}$ QPOs and scaling (\ref{equation:bestfit}) could provide the basis for a method intended to determine the mass of BH sources such as active galactic nuclei (AGN) and ultraluminuos X-ray (ULX) sources. It has been argued that a confirmation of ${3\!:\!2}$ QPOs in other than stellar mass BH sources  could be of fundamental importance for the BH accretion theory \cite[][]{abr-etal:2004:ApJ:,tor:2005:AA:,tor:2005:AN:}.} It was furthermore discussed that within a wide range of BH masses $M,$ both the rotational parameter $a\equiv\mathrm{c}J/\mathrm{G}M^2$ and specific details of a given orbital QPO model are of secondary importance, and that the observed frequencies can be used for to estimate $M$.\footnote{In this paper we assume Kerr BH space times.}

\begin{figure*}[t]
\begin{center}
\includegraphics[width=.8\hsize]{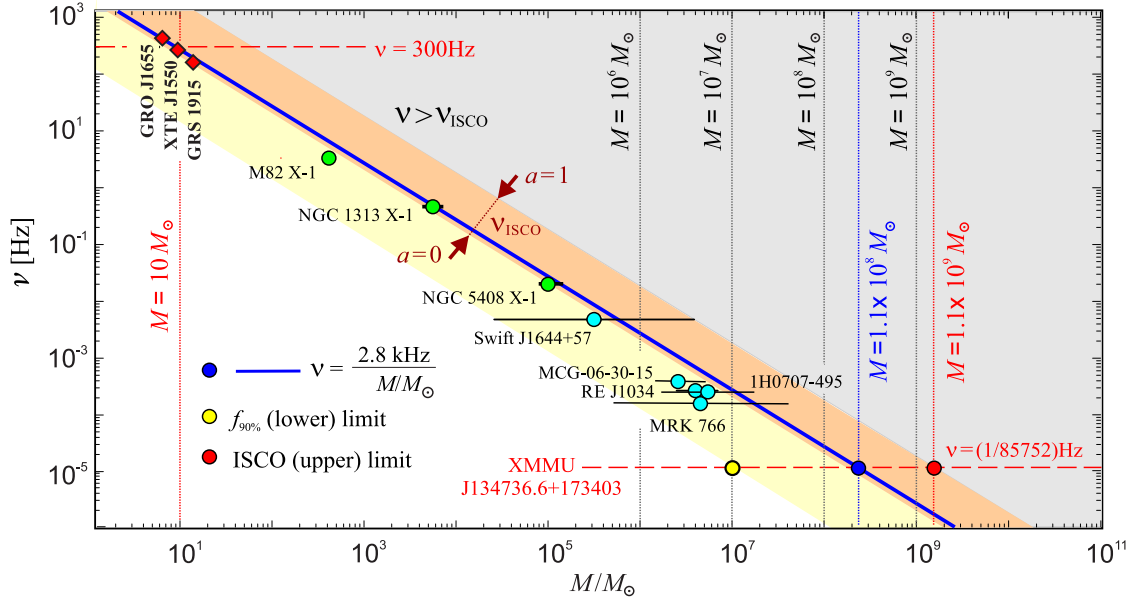}
\end{center}
\vspace{-2ex}

\noindent
\caption{Large scaling of BH {HF} QPO frequencies vs. Keplerian frequencies in the accretion disk. The upper left corner of the plot corresponds to Galactic microquasar BHs, while the lower right corner corresponds to supermassive BHs. The light orange region denotes the ISCO frequencies in the range of $a\in[0,~1]$. The  gray area indicates frequencies that are higher than the ISCO frequency. The light yellow area denotes Keplerian frequencies in the inner part of the (thin) disk that radiates more than 90\% of the whole disk luminosity ($a\geq0$, see Section~\ref{section:models}). {The green circles denote intermediate-mass BH sources whose mass estimate is based either fully or in a large part on the observations of HF QPOs. For some sources, the error bars are  smaller than the displayed symbols.}}
\label{figure:scaling}
\end{figure*}

A full decade after these results have been published, \cite{zho-etal:2015} completed a survey pointing to a further observational evidence for the large-scale validity of the $1/M$ scaling of the BH QPO frequencies (\ref{equation:bestfit}). Even more important findings that support this evidence have been accumulated recently. At present, several sources that span the range of a few orders of magnitude seem to follow the relation. This is illustrated in Figure~\ref{figure:scaling}, and further details along with several references can be found in \cite{zho-etal:2015}, {\cite{pas-etal:2014,pas-etal:2015}}, {\cite{cze-etal:2016}}, \cite{zha-etal:2017,zha-etal:2018}, \cite{gup-etal:2018}, and also \cite{smi-etal:2018}. 
We here discuss the implications of the latest QPO observations and mainly focus on the importance of QPOs that have recently been found in the X-ray source \xmmu~ by \cite{car-etal:2018}.

%

\section{Upper bounds on AGN mass based on HF QPOs}

\cite{car-etal:2018} carried out  a systematic study of Chandra, Swift, and XMM-Newton observations of the X-ray source \xmmu,~ which has been previously found to be spatially coincident with a pair of galaxies including the Seyfert 2 galaxy SDSS J134736.39+173404.6. Using the Chandra observation from 2008, \cite{car-etal:2018} accurately evaluated the position of this X-ray source and showed that its coordinates coincide with the position of the Seyfert 2 galaxy. Within the analysis of a set of 29 Swift observations conducted between February 6 and May 23, 2008, they discovered strong twin-peak quasi-periodic oscillations with periods of 23.82$\pm$0.07 h and 71.44$\pm$0.57 h.

This measurement clearly indicates that $R=3.00\pm{0.04}$, and it provides very strong evidence for frequency-commensurable QPOs in an AGN BH. The frequency commensurability is crucial for the nonlinear  resonance  models  discussed  by  Abramowicz,  Kluzniak and collaborators \citep[e.g.,][]{abr-klu:2001,reb:2004:PASJ:,hor:2008:AA:,tor-etal:2005}. Motivated by the observed $3:1$ frequency ratio, \cite{car-etal:2018} investigated the mass-spin relations expected for this source in detail from various resonances between disk-oscillation modes that have previously been discussed in the context of QPOs in Galactic microquasars. We make several remarks regarding these models in Sections~\ref{section:models} and \ref{section:resonances}, where other QPO models are considered as well. In what follows, we explore general implications of the observed rapid variability that are valid for a large class of orbital QPO models.


The Keplerian frequency of matter orbiting a BH monotonically increases as the orbital radius {$r$} decreases to the inner edge of the accretion disk. The location of the inner edge depends on the radiative efficiency of the disk. For a very high efficiency, the disk terminates at the marginally stable circular orbit (thin disks), $\rISCO$ (often called innermost stable circular orbit, or ISCO), while for a very low efficiency, it terminates at the marginally bound orbit (ion tori, thin disks, and {advection dominated accretion flows - ADAFs}), $\rRISCO$. The Keplerian orbital frequency at these orbits for a Schwarzschild BH ($a=0$) scales with BH mass as \citep[e.g.,][]{bar-etal:1972} 
\begin{equation}
\label{equation:ISCO}
\nuISCO=\frac{2.20\mathrm{kHz}}{M^{*}},\quad \nuRISCO=\frac{4.04\mathrm{kHz}}{M^{*}}.
\end{equation}
For rotating BHs (corrotating disks), these frequencies are higher, and for {maximally rotating} Kerr BH ($a=1$), we may write
\begin{equation}
\nuISCO=\nuRISCO=\frac{16.2\mathrm{kHz}}{M^{*}}.
\end{equation}

In Figure~\ref{figure:scaling} we present relations that determine the highest orbital frequencies. We also mark here the higher QPO frequency observed in \xmmu. Postulating that this frequency corresponds to the Keplerian frequency inside the disk, we can find that the mass of the source should not be higher than $M\doteq1.1\times 10^{9}M_{\sun}$.

\section{Application of particular orbital models of QPOs and lower bounds on AGN mass}
\label{section:models}

\begin{figure*}[t]
\begin{center}
a)\hfill ~~~~~~~~~~~~~~~b) \hfill $\phantom{c)}$
\includegraphics[width=1.\hsize]{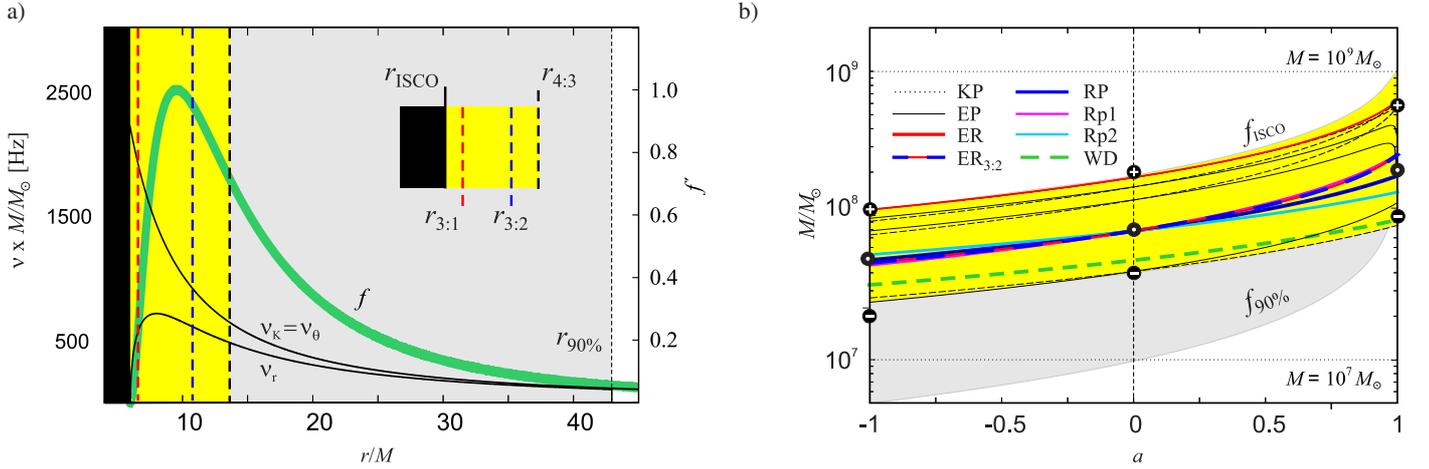}
\end{center}
\vspace{-3ex}

\noindent
\caption{a) Radial profiles of the normalized disk flux  (thin relativistic disk) and the Keplerian and epicyclic frequencies of geodesic orbital motion calculated for a Schwarzschild BH. The gray region between $\rISCO$ and $r_{90\%}$ emphasizes the radial part of the disk that radiates more than 90\% of the whole disk luminosity. Several particular resonant orbits are denoted. Indexes i and j in $r_{i:j}$ are chosen such that $i/j=\nuv/\nur$. b) Mass-spin relations inferred for \xmmu~ from the observed QPO frequencies and their models in Kerr spacetimes. The solid and dashed black lines are drawn for the set of resonance models considered by \cite{car-etal:2018}. The red line denotes the prediction of the ER model, while the blue line corresponds to the RP model. The dashed red-blue line denotes the consideration of the ER$_{3:2}$ model. The gray region corresponds to the association of the QPO frequency with the Keplerian frequency at orbits inside a part of the disk that radiates more than 90\% of the whole disk luminosity. The lower boundary of the gray region corresponds to  $r_{90\%}$ , while the upper boundary corresponds to $\rISCO$. The yellow region emphasizes the range covered by the assumed QPO models. The {nine} spots denote boundaries of relation (\ref{relation:SGO}) for {$a\in[-1,~1]$}.}
\label{figure:m-a}

\end{figure*}

A large group of models, to which we refer as standard geodesic orbital (SGO) models, assume that the observed frequencies, $\nuL$ and $\nuU$, are equal to fundamental frequencies of (geodesic) orbital motion, that is, the Keplerian frequency defined at a circular orbit above {ISCO, $r\geq\rISCO$,} and the radial and vertical epicyclic frequency, or to their linear combinations (including the periastron and Lense-Thirring precession frequency). The three frequencies $\nuK$, $\nur$ , and $\nuv$ can be written in a general form 
\begin{equation}
\label{equation:uni}
\nu_{\mathrm{i}} = \frac{1}{T_{\mathrm{i}}}= \frac{c^3}{2\pi  G}\frac{1}{M}{\mathcal{F}}_{\mathrm{i}}(x,a) \leq \nuISCO(M,a),
\end{equation}
where $x=r/r_{G}\geq\rISCO/r_{G}$, $r_{G}=GM/c^2${, and $\mathcal{F}_{\mathrm{i}}$ is a specific function that for a fixed BH spin depends only on $x$}. The $1/M$ term here provides a physical explanation of the empiric scaling (\ref{equation:bestfit}). We illustrate the behavior of Keplerian and epicyclic frequencies for the simple case of a Schwarzschild BH (a=0) in Figure~{\ref{figure:m-a}}a. We also include in this figure an illustration of the behavior of the normalized disk flux $f=f/f_{\mathrm{max}}$ calculated for relativistic thin disks \citep{pag-don:1974}.

\begin{table}
\caption{{Frequency relations corresponding to individual QPO models investigated in \cite{tor-etal:2011:AA}. The relations are expressed in terms of three fundamental frequencies of perturbed circular geodesic motion.}}
\label{table:models}
\renewcommand{\arraystretch}{1.}
\begin{center}
\begin{tabular}{lll}
    \hline
  \hline
     \textbf{Model} & \multicolumn{2}{c}{\textbf{Relations}} \\
    \hline \hline
$\mathbf{RP}$ & $\nu_{\mathrm{L}} =\nu_{\mathrm{K}}-\nu_{\mathrm{r}}$ & $\nu_{\mathrm{U}} =\nu_{\mathrm{K}}$  \\
$\mathbf{TD}^{*}$ & $\nu_{\mathrm{L}}=\nu_{\mathrm{K}}$ & $\nu_{\mathrm{U}}=\nu_{\mathrm{K}}+\nu_{\mathrm{r}}$ \\
\hline
$\mathbf{WD}$ & $\nu_{\mathrm{L}}=2\left(\nu_{\mathrm{K}}-\nu_{\mathrm{r}}\right)$ & $\nu_{\mathrm{U}}=2\nu_{\mathrm{K}}-\nu_{\mathrm{r}}$ \\
$\mathbf{ER}^{\nabla}$ & $\nu_{\mathrm{L}}=\nu_{\mathrm{r}}$ & $\nu_{\mathrm{U}}=\nu_{\theta}$ \\
$\mathbf{KeP}^{\nabla}$ & $\nu_{\mathrm{L}}=\nu_{\mathrm{r}}$ & $\nu_{\mathrm{U}}=\nu_{\mathrm{K}}$ \\
$\mathbf{RP1}$ &  $\nu_{\mathrm{L}}=\nu_{\mathrm{K}}-\nu_{\mathrm{r}}$ &  $\nu_{\mathrm{U}}= \nu_{\theta}$ \\
$\mathbf{RP2}$ &  $\nu_{\mathrm{L}}=\nu_{\mathrm{K}}-\nu_{\mathrm{r}}$ &  $\nu_{\mathrm{U}}= 2\nu_{\mathrm{K}}-\nu_{\theta}$ \\
\hline \hline
\end{tabular}
\end{center}

{\scriptsize{$^{*}$Model is not consistent with the observed $3:1$ frequency ratio.\\ $^{\nabla}$Model overlaps with the consideration of \cite{car-etal:2018}.}}
`

\end{table}

The family of SGO models includes various physical concepts that hold the assumption that the QPO excitation radii are located within the most luminuos accretion region ($r\!\ll\!r_{90\%}$, see Figure~{\ref{figure:m-a}}a), usually below $20r_{\mathrm{G}}$. Several models assume, for instance, that QPOs are produced by a local motion of accreted inhomogeneities such as blobs or vortices. This subset of SGO models includes the so-called relativistic precession model (RP) or tidal disruption (TD) model \citep[][]{abr-etal:1992,ste-vie:1998,ste-vie:1999,cad-etal:2008,kos-etal:2009,bak-etal:2014,kar-etal:2017,ger-etal:2017}. Another possibility is to relate the QPOs to a collective motion of the accreted matter, in particular, to some accretion disk oscillatory modes \cite[][]{wag-etal:2001,rez-etal:2003,abr-etal:2006,Ingram+Done:2010,fragile-blaes:2016}. Specific models of this type often deal with oscillations in a slender accretion torus and assume some kind of resonance between these oscillation modes.

A general idea {considering} such resonances has originated in \citet{abr-klu:2001} \citep[see also][]{ali-gal:1981} and was later continued in the context of BH QPOs and spin extensively discussed in several subsequent works covering both parametric and forced resonances \citep[][]{tor-etal:2005,tor:2005:AA:,tor:2005:AN:,tor-etal:2011:AA,stu-etal:2013:}. Various combinations of both axisymmetric and non-axisymmetric epicyclic modes (EP class of models), as well as combinations of these modes and the Keplerian oscillation (KP class of models), have been considered. An often-quoted model is represented by the $3\!:\!2$ epicyclic resonance model, or by a similar model that consideres the Keplerian frequency instead of the vertical epicyclic frequency (ER and KeP models).


The mass-spin relations predicted for Galactic microquasars by several QPO models have been investigated in a series of works \citep{tor-etal:2011:AA,kot-etal:2014,kot-etal:2017}. Along with the TD, RP, ER, and KeP model, these include the specific diskoseismic model of \cite{kato:2001:PASJ:,kat:2007:PASJ:,kat:2008:PASJ:} and two models introduced by \cite{bur:2005} and \cite{tor-etal:2007,Tor-etal:2010} that were motivated by the observations of NS QPOs (the RP1 and RP2 models). We recall in Table1 how  the QPO frequencies are identified with orbital frequencies of geodesic motion.

We calculated the mass-spin relations for \xmmu~ based on the models listed in  Table~\ref{table:models}. The obtained results are presented in Figure~\ref{figure:m-a}b along with all mass-spin relations relevant to the models considered by \cite{car-etal:2018}. The mass predicted by these models is always below the ISCO limit $M\doteq1.1\times 10^{9}M_{\sun}$. {It is nevertheless close to this limit (see the horizontal range of Figure~\ref{figure:scaling}),} and there is $10^{7}M_{\sun}<M< 10^{9}M_{\sun}$, which is generic for the proposed SGO models when $a\geq0$ because they involve QPO excitation radii $r<r_{90\%}$. {In Figure~{\ref{figure:m-a}}b we illustrate these findings and plot functions $f_{90\%}\equiv M(r_{90\%})$ and $f_{\ISCO}\equiv M(\rISCO)$ that correspond to an identification of the observed QPO frequency with the appropriate Keplerian frequency.}

\section{Implications for models assuming parametric resonances}
\label{section:resonances}

\citet{klu-abr:2002} have suggested that the lower and upper kHz QPOs may be identified with the radial and vertical axisymmetric epicyclic oscillations of an accretion disk. In their scenario, the resonance arises as a result of nonlinear coupling between the oscillation modes.  The radial mode supplies energy to the vertical mode by means of the parametric resonance. As the parametric resonance occurs when it is $\nu_r/\nu_\theta = 2/n$ (with $n$ being an integer number) and $\nu_r\leq\nu_\theta$, their model naturally explains the $3\!:\!2$ frequency ratio that is often observed in the QPO sources. A quantitative analysis of this model has been presented by \citet{reb:2004:PASJ:} and \citet{hor:2008:AA:}. The latter work is devoted to a detailed study of the nature of nonlinear coupling between disk oscillation modes and extends the previous analysis to non-axisymmetric epicyclic modes. Consideration of these modes implies frequency identifications of the lower and upper QPO that involve combinations of both Keplerian and epicyclic frequencies because they oscillate with frequencies $m\nu_\mathrm{K} \pm \nu_{\mathrm{r},\theta}$, where $m$ is an integer azimuthal wavenumber. 

\subsection{Restrictions on resonant interactions}
\balance

Formulae of the Ep, RP1, and RP2 models from Table~\ref{table:models} combine various combinations of the radial and vertical epicylic modes with azimuthal wavenumbers $(m_\mathrm{r}, m_\theta) = (0,0)$, $(1,0),$ and $(1,2)$, respectively. The symmetry properties of the accretion flow strongly limit the resonant interaction between the modes. In particular, when the unperturbed flow is both axially symmetric and symmetric with respect to the equatorial plane, possible resonances up to the fourth order occur only when $\nu_\theta/\nu_r = 1/2, 1, 1/4,$ and $3/2$, and the same condition has also to be satisfied by the azimuthal wavenumbers, that is, $m_\theta/m_r = 1/2$, 1, $1/4$ or $3/2$.  In this context, in the case of the $3\!:\!2$ observed QPO frequency ratio, the idea of parametric resonance is consistent only with the frequency identification provided by the ER model \citep{hor:2008:AA:}. Even this model nevertheless fails in explaining the 3:1 frequency ratio observed in XMMUJ134736.6+173403.

\subsection{$3\!:\!2$ resonance}

A resolution of the problem may be connected to a nonlinear nature of both the flow oscillations and the modulation mechanism through which the oscillations manifest themselves in the observed light curves. The observed power spectra often contain higher harmonics and subharmonics of the fundamental mode frequencies. The 3:1 frequency ratio then still may be explained by the $3\!:\!2$ resonance provided that the first-order harmonics of the upper fundamental mode, or a subharmonic of the lower fundamental mode, is observed. We illustrate the underlying mass-spin relation in Figure~\ref{figure:m-a}. It is considered that $\nuU$ equals $\nuv$ and $\nuL$ equals $\nuv-\nur$ (ER$_{3:2}$ model).

\section{Discussion and conclusions}
\label{section:conclusions}

If the observed BH HF QPOs can be described by a {model} similar to some of the SGO models, \xmmu~  likely represents the most massive BH source with commensurable QPO frequencies. Based on the mass estimation implied by the ISCO limit, we can establish $M\approx 10^9M_{\sun}$ as the current observational upper limit on an AGN BH mass inferred from QPOs. This limit is valid as long as the real oscillatory frequencies of the disk are not much higher than the Keplerian frequency. This case is nevertheless not very likely to occur \citep[e.g.,][]{str-sram:2009}. We note that consideration of miscellaneous combinational frequencies within the QPO models may lead to obtaining observable frequencies {that are (although not much) higher  than} the Keplerian limit.

\subsection{{Local versus global models, resonances}}

{{The problem of determination of the right qpo model exceeds the scope of this Letter.} {It is nevertheless worthwhile to note that HF QPOs commonly found in Seyefert 1 and Seyefert 2 AGNs pose a challenge for physical concepts that have been proposed to explain modulation of the observed flux within the QPO models.}
}

{Seyfert 1 AGNs are expected to have accretion disks with small inclination angles (i.e., disks that are nearly parallel to the observer's sky). Models that include orbiting accreted inhomogeneities assume a rather small extent of the modulated region of the disk, which implies a low ratio of the modulated to total disk flux. This ratio often approaches zero for the nearly face-on view \citep[e.g,][]{sch-ber:2004}. Either somewhat non-equatorial orbits or a very powerful enhancement of the inhomogeneities are therefore required in order to explain the observed source variability (e.g., tidal disruption events). In principle, models that introduce a collective motion of the accreted fluid allow for a high spatial extent of the modulated accretion region \citep[e.g.,][]{wag-etal:2001,bur-etal:2004,sch-rez:2006,Ingram+Done:2010,bak-etal:2014}. In this sense, this group of models that include the diskoseismic models and concepts that incorporate oscillating tori or hot inner accretion flow are favored.}

{Regarding the latter class of models, we} suggest that it will be crucial for the falsification of models that {are based on} parametric resonances {between disk oscillation modes} whether only single QPO with a frequency that completes the harmonic sequence $3\!:\!2\!:\!1$ is found in this source, or if a new different pair of QPOs with frequencies in the $3:\!2$ ratio is found. The former case would agree with the prediction of the $3\!:\!2$ epicyclic resonance (ER$_{3:2}$) model, which explains the observed 3:1 frequency ratio by means of the combinational frequencies.
For {both corotating and counterrotating} disks, the corresponding mass-spin relation shown in Figure~\ref{figure:m-a} (the dashed red-blue line) can be matched by a simple approximative quadratic relation, ${M\approx(5a^2 + 8a + 8)\times 10^{7}M_{\sun}}$.

\subsection{{Quantification of results common to SGO models}}

Figure~\ref{figure:m-a}b shows that the ER$_{3:2}$ model prediction coincides for $a=0$ with the predictions of several other models such as the RP model. On a large scale of $M,$ all the considered models provide similar mass-spin relations. Following the quadratic approximation, we may write
\begin{equation}
\label{relation:SGO}
{M\approx(5^{+11}_{-3}a^2 + 8^{+17}_{-4}a + 8^{+12}_{-4} )\times 10^{7}M_{\sun}},
\end{equation}
where the upper limit corresponds to models that require a high BH mass (e.g., the ER model) and the lower limit corresponds to models that require a low BH mass (e.g., the WD model). The boundaries of relation (\ref{relation:SGO}) for {$a\in[-1,~1]$} are denoted by spots in Figure~\ref{figure:m-a}b. This figure shows  that relation (\ref{relation:SGO}) describes the predictions of the whole group of the considered SGO models well.

\section*{Acknowledgments}
\renewcommand{\baselinestretch}{.95}
{We would like to acknowledge the Czech Science Foundation grant No. 17-16287S, the INTER-EXCELLENCE project No. LTI17018 that supports the collaboration between the Silesian University in Opava and the Astronomical Institute in Prague, and internal grants of the Silesian University, SGS/14,15/2016. {We thank the anonymous referee for their comments and suggestions that have greatly helped us to improve the paper.} 
}

\bibliographystyle{aa}
\renewcommand{\baselinestretch}{.87}
\bibliography{reference}

\begin{thebibliography}{61}
\expandafter\ifx\csname natexlab\endcsname\relax\def\natexlab#1{#1}\fi

\bibitem[{{Abramowicz} {et~al.}(2003){Abramowicz}, {Almergren}, {Klu{\'z}niak},
  \& {Thampan}}]{abr-etal:2003a}
{Abramowicz}, M.~A., {Almergren}, G.~J.~E., {Klu{\'z}niak}, W., \& {Thampan},
  A.~V. 2003, ArXiv General Relativity and Quantum Cosmology e-prints

\bibitem[{{Abramowicz} {et~al.}(2006){Abramowicz}, {Blaes}, {Hor{\'a}k},
  {Klu{\'z}niak}, \& {Rebusco}}]{abr-etal:2006}
{Abramowicz}, M.~A., {Blaes}, O.~M., {Hor{\'a}k}, J., {Klu{\'z}niak}, W., \&
  {Rebusco}, P. 2006, Classical and Quantum Gravity, 23, 1689

\bibitem[{{Abramowicz} \& {Klu{\'z}niak}(2001)}]{abr-klu:2001}
{Abramowicz}, M.~A. \& {Klu{\'z}niak}, W. 2001, A\&A, 374, L19

\bibitem[{{Abramowicz} {et~al.}(2004){Abramowicz}, {Klu{\'z}niak},
  {McClintock}, \& {Remillard}}]{abr-etal:2004:ApJ:}
{Abramowicz}, M.~A., {Klu{\'z}niak}, W., {McClintock}, J.~E., \& {Remillard},
  R.~A. 2004, APJL, 609, L63

\bibitem[{{Abramowicz} {et~al.}(1992){Abramowicz}, {Lanza}, {Spiegel}, \&
  {Szuszkiewicz}}]{abr-etal:1992}
{Abramowicz}, M.~A., {Lanza}, A., {Spiegel}, E.~A., \& {Szuszkiewicz}, E. 1992,
  Nature, 356, 41

\bibitem[{{Aliev} \& {Galtsov}(1981)}]{ali-gal:1981}
{Aliev}, A.~N. \& {Galtsov}, D.~V. 1981, Gen. Relativ. and Gravitation, 13, 899

\bibitem[{{Bakala} {et~al.}(2014){Bakala}, {T{\"o}r{\"o}k}, {Karas}, {Dov{\v
  c}iak}, {Wildner}, {Wzientek}, {{\v S}r{\'a}mkov{\'a}}, {Abramowicz},
  {Goluchov{\'a}}, {Mazur}, \& {Vincent}}]{bak-etal:2014}
{Bakala}, P., {T{\"o}r{\"o}k}, G., {Karas}, V., {et~al.} 2014, MNRAS, 439, 1933

\bibitem[{{Bardeen} {et~al.}(1972){Bardeen}, {Press}, \&
  {Teukolsky}}]{bar-etal:1972}
{Bardeen}, J.~M., {Press}, W.~H., \& {Teukolsky}, S.~A. 1972, APJ, 178, 347

\bibitem[{{Barret} \& {Boutelier}(2008)}]{bar-bou:2008}
{Barret}, D. \& {Boutelier}, M. 2008, New Astron. Rev., 51, 835

\bibitem[{{Belloni} {et~al.}(2007){Belloni}, {Homan}, {Motta}, {Ratti}, \&
  {M{\'e}ndez}}]{bel-etal:2007}
{Belloni}, T., {Homan}, J., {Motta}, S., {Ratti}, E., \& {M{\'e}ndez}, M. 2007,
  \mnras, 379, 247

\bibitem[{{Belloni} {et~al.}(2005){Belloni}, {M{\'e}ndez}, \&
  {Homan}}]{bel-etal:2005}
{Belloni}, T., {M{\'e}ndez}, M., \& {Homan}, J. 2005, A\&A, 437, 209

\bibitem[{{Belloni} \& {Altamirano}(2013)}]{bel-alt:2013}
{Belloni}, T.~M. \& {Altamirano}, D. 2013, \mnras, 432, 10

\bibitem[{{Belloni} {et~al.}(2012){Belloni}, {Sanna}, \&
  {M{\'e}ndez}}]{bel-etal:2012}
{Belloni}, T.~M., {Sanna}, A., \& {M{\'e}ndez}, M. 2012, \mnras, 426, 1701

\bibitem[{{Boutelier} {et~al.}(2010){Boutelier}, {Barret}, {Lin}, \&
  {T{\"o}r{\"o}k}}]{bou-etal:2010}
{Boutelier}, M., {Barret}, D., {Lin}, Y., \& {T{\"o}r{\"o}k}, G. 2010, MNRAS,
  401, 1290

\bibitem[{{Bursa}(2005)}]{bur:2005}
{Bursa}, M. 2005, in RAGtime 6/7: Workshops on black holes and neutron stars,
  ed. S.~{Hled{\'{\i}}k} \& Z.~{Stuchl{\'{\i}}k}, 39--45

\bibitem[{{Bursa} {et~al.}(2004){Bursa}, {Abramowicz}, {Karas}, \&
  {Klu{\'z}niak}}]{bur-etal:2004}
{Bursa}, M., {Abramowicz}, M.~A., {Karas}, V., \& {Klu{\'z}niak}, W. 2004,
  \apjl, 617, L45

\bibitem[{{Carpano} \& {Jin}(2018)}]{car-etal:2018}
{Carpano}, S. \& {Jin}, C. 2018, MNRAS, 477, 3178

\bibitem[{{Czerny} {et~al.}(2016){Czerny}, {You}, {Kurcz},
  {{\'S}redzi{\'n}ska}, {Hryniewicz}, {Niko{\l}ajuk}, {Krupa}, {Wang}, {Hu}, \&
  {{\.Z}ycki}}]{cze-etal:2016}
{Czerny}, B., {You}, B., {Kurcz}, A., {et~al.} 2016, \aap, 594, A102

\bibitem[{{Fragile} {et~al.}(2016){Fragile}, {Straub}, \&
  {Blaes}}]{fragile-blaes:2016}
{Fragile}, P.~C., {Straub}, O., \& {Blaes}, O. 2016, MNRAS, 461, 1356

\bibitem[{{German{\`a}}(2017)}]{ger-etal:2017}
{German{\`a}}, C. 2017, PRD, 96, 103015

\bibitem[{{Gilfanov} {et~al.}(2000){Gilfanov}, {Churazov}, \&
  {Revnivtsev}}]{gilf-etal:2000}
{Gilfanov}, M., {Churazov}, E., \& {Revnivtsev}, M. 2000, MNRAS, 316, 923

\bibitem[{{Gupta} {et~al.}(2018){Gupta}, {Tripathi}, {Wiita}, {Gu}, {Bambi}, \&
  {Ho}}]{gup-etal:2018}
{Gupta}, A.~C., {Tripathi}, A., {Wiita}, P.~J., {et~al.} 2018, A\&A, 616, L6

\bibitem[{{Hor{\'a}k}(2008)}]{hor:2008:AA:}
{Hor{\'a}k}, J. 2008, A\&A, 486, 1

\bibitem[{{Ingram} \& {Done}(2010)}]{Ingram+Done:2010}
{Ingram}, A. \& {Done}, C. 2010, MNRAS, 405, 2447

\bibitem[{{Karssen} {et~al.}(2017){Karssen}, {Bursa}, {Eckart}, {Valencia-S},
  {Dov{\v c}iak}, {Karas}, \& {Hor{\'a}k}}]{kar-etal:2017}
{Karssen}, G.~D., {Bursa}, M., {Eckart}, A., {et~al.} 2017, MNRAS, 472, 4422

\bibitem[{{Kato}(2001)}]{kato:2001:PASJ:}
{Kato}, S. 2001, PASJ, 53, 1

\bibitem[{{Kato}(2007)}]{kat:2007:PASJ:}
{Kato}, S. 2007, PASJ, 59, 451

\bibitem[{{Kato}(2008)}]{kat:2008:PASJ:}
{Kato}, S. 2008, PASJ, 60, 111

\bibitem[{{Klu{\'z}niak} \& {Abramowicz}(2002)}]{klu-abr:2002}
{Klu{\'z}niak}, W. \& {Abramowicz}, M.~A. 2002, ArXiv Astrophysics e-prints

\bibitem[{{Kosti{\'c}} {et~al.}(2009){Kosti{\'c}}, {{\v C}ade{\v z}},
  {Calvani}, \& {Gomboc}}]{kos-etal:2009}
{Kosti{\'c}}, U., {{\v C}ade{\v z}}, A., {Calvani}, M., \& {Gomboc}, A. 2009,
  A\&A, 496, 307

\bibitem[{{Kotrlov{\'a}} {et~al.}(2014){Kotrlov{\'a}}, {T{\"o}r{\"o}k}, {{\v
  S}r{\'a}mkov{\'a}}, \& {Stuchl{\'{\i}}k}}]{kot-etal:2014}
{Kotrlov{\'a}}, A., {T{\"o}r{\"o}k}, G., {{\v S}r{\'a}mkov{\'a}}, E., \&
  {Stuchl{\'{\i}}k}, Z. 2014, A\&A, 572, A79

\bibitem[{{Kotrlov{\'a}} {et~al.}(2017){Kotrlov{\'a}}, {{\v S}r{\'a}mkov{\'a}},
  {T{\"o}r{\"o}k}, {Stuchl{\'{\i}}k}, \& {Goluchov{\'a}}}]{kot-etal:2017}
{Kotrlov{\'a}}, A., {{\v S}r{\'a}mkov{\'a}}, E., {T{\"o}r{\"o}k}, G.,
  {Stuchl{\'{\i}}k}, Z., \& {Goluchov{\'a}}, K. 2017, A\&A, 607, A69

\bibitem[{{McClintock} \& {Remillard}(2006)}]{mcc-rem:2006}
{McClintock}, J.~E. \& {Remillard}, R.~A. 2006, {Black hole binaries}
  (Cambridge University Press), 157--213

\bibitem[{{Page} \& {Thorne}(1974)}]{pag-don:1974}
{Page}, D.~N. \& {Thorne}, K.~S. 1974, APJ, 191, 499

\bibitem[{{Pasham} {et~al.}(2015){Pasham}, {Cenko}, {Zoghbi}, {Mushotzky},
  {Miller}, \& {Tombesi}}]{pas-etal:2015}
{Pasham}, D.~R., {Cenko}, S.~B., {Zoghbi}, A., {et~al.} 2015, APJL, 811, L11

\bibitem[{{Pasham} {et~al.}(2014){Pasham}, {Strohmayer}, \&
  {Mushotzky}}]{pas-etal:2014}
{Pasham}, D.~R., {Strohmayer}, T.~E., \& {Mushotzky}, R.~F. 2014, \nat, 513, 74

\bibitem[{{Rebusco}(2004)}]{reb:2004:PASJ:}
{Rebusco}, P. 2004, PASJ, 56, 553

\bibitem[{{Remillard} {et~al.}(2002){Remillard}, {Muno}, {McClintock}, \&
  {Orosz}}]{rem-etal:2002}
{Remillard}, R.~A., {Muno}, M.~P., {McClintock}, J.~E., \& {Orosz}, J.~A. 2002,
  \apj, 580, 1030

\bibitem[{{Rezzolla} {et~al.}(2003){Rezzolla}, {Yoshida}, \&
  {Zanotti}}]{rez-etal:2003}
{Rezzolla}, L., {Yoshida}, S., \& {Zanotti}, O. 2003, MNRAS, 344, 978

\bibitem[{{Schnittman} \& {Bertschinger}(2004)}]{sch-ber:2004}
{Schnittman}, J.~D. \& {Bertschinger}, E. 2004, \apj, 606, 1098

\bibitem[{{Schnittman} \& {Rezzolla}(2006)}]{sch-rez:2006}
{Schnittman}, J.~D. \& {Rezzolla}, L. 2006, \apjl, 637, L113

\bibitem[{{Smith} {et~al.}(2018){Smith}, {Mushotzky}, {Boyd}, \&
  {Wagoner}}]{smi-etal:2018}
{Smith}, K.~L., {Mushotzky}, R.~F., {Boyd}, P.~T., \& {Wagoner}, R.~V. 2018,
  APJL, 860, L10

\bibitem[{{Stella} \& {Vietri}(1998)}]{ste-vie:1998}
{Stella}, L. \& {Vietri}, M. 1998, AJ Lett., 492, L59

\bibitem[{{Stella} {et~al.}(1999){Stella}, {Vietri}, \&
  {Morsink}}]{ste-vie:1999}
{Stella}, L., {Vietri}, M., \& {Morsink}, S.~M. 1999, AJ Lett., 524, L63

\bibitem[{{Straub} \& {{\v S}r{\'a}mkov{\'a}}(2009)}]{str-sram:2009}
{Straub}, O. \& {{\v S}r{\'a}mkov{\'a}}, E. 2009, Classical and Quantum
  Gravity, 26, 055011

\bibitem[{{Stuchl{\'{\i}}k} {et~al.}(2013){Stuchl{\'{\i}}k}, {Kotrlov{\'a}}, \&
  {T{\"o}r{\"o}k}}]{stu-etal:2013:}
{Stuchl{\'{\i}}k}, Z., {Kotrlov{\'a}}, A., \& {T{\"o}r{\"o}k}, G. 2013, A\&A,
  552, A10

\bibitem[{{T{\"o}r{\"o}k}(2005{\natexlab{a}})}]{tor:2005:AA:}
{T{\"o}r{\"o}k}, G. 2005{\natexlab{a}}, APJ, 440, 1

\bibitem[{{T{\"o}r{\"o}k}(2005{\natexlab{b}})}]{tor:2005:AN:}
{T{\"o}r{\"o}k}, G. 2005{\natexlab{b}}, Astronomische Nachrichten, 326, 856

\bibitem[{{T{\"o}r{\"o}k} {et~al.}(2008{\natexlab{a}}){T{\"o}r{\"o}k},
  {Abramowicz}, {Bakala}, {Bursa}, {Hor{\'a}k}, {Kluzniak}, {Rebusco}, \&
  {Stuchlik}}]{tor-etal:2008b}
{T{\"o}r{\"o}k}, G., {Abramowicz}, M.~A., {Bakala}, P., {et~al.}
  2008{\natexlab{a}}, Acta Astron., 58, 15

\bibitem[{{T{\"o}r{\"o}k} {et~al.}(2005){T{\"o}r{\"o}k}, {Abramowicz},
  {Klu{\'z}niak}, \& {Stuchl{\'{\i}}k}}]{tor-etal:2005}
{T{\"o}r{\"o}k}, G., {Abramowicz}, M.~A., {Klu{\'z}niak}, W., \&
  {Stuchl{\'{\i}}k}, Z. 2005, A\&A, 436, 1

\bibitem[{{T{\"o}r{\"o}k} {et~al.}(2008{\natexlab{b}}){T{\"o}r{\"o}k},
  {Bakala}, {Stuchlik}, \& {\v{C}ech}}]{tor-etal:2008a}
{T{\"o}r{\"o}k}, G., {Bakala}, P., {Stuchlik}, Z., \& {\v{C}ech}, P.
  2008{\natexlab{b}}, Acta Astron., 58, 1

\bibitem[{{T{\"o}r{\"o}k} {et~al.}(2010){T{\"o}r{\"o}k}, {Bakala}, {{\v
  S}r{\'a}mkov{\'a}}, {Stuchl{\'{\i}}k}, \& {Urbanec}}]{Tor-etal:2010}
{T{\"o}r{\"o}k}, G., {Bakala}, P., {{\v S}r{\'a}mkov{\'a}}, E.,
  {Stuchl{\'{\i}}k}, Z., \& {Urbanec}, M. 2010, AJ, 714, 748

\bibitem[{{T{\"o}r{\"o}k} {et~al.}(2011){T{\"o}r{\"o}k}, {Kotrlov{\'a}}, {{\v
  S}r{\'a}mkov{\'a}}, \& {Stuchl{\'{\i}}k}}]{tor-etal:2011:AA}
{T{\"o}r{\"o}k}, G., {Kotrlov{\'a}}, A., {{\v S}r{\'a}mkov{\'a}}, E., \&
  {Stuchl{\'{\i}}k}, Z. 2011, A\&A, 531, A59

\bibitem[{{T{\"o}r{\"o}k} {et~al.}(2007){T{\"o}r{\"o}k}, {Stuchl{\'{\i}}k}, \&
  {Bakala}}]{tor-etal:2007}
{T{\"o}r{\"o}k}, G., {Stuchl{\'{\i}}k}, Z., \& {Bakala}, P. 2007, Cent.
  European J. of Phys., 5, 457

\bibitem[{{{\v C}ade{\v z}} {et~al.}(2008){{\v C}ade{\v z}}, {Calvani}, \&
  {Kosti{\'c}}}]{cad-etal:2008}
{{\v C}ade{\v z}}, A., {Calvani}, M., \& {Kosti{\'c}}, U. 2008, A\&A, 487, 527

\bibitem[{{van der Klis}(2006)}]{kli:2006}
{van der Klis}, M. 2006, {Rapid X-ray Variability} (Cambridge University
  Press), 39--112

\bibitem[{{Varniere} \& {Rodriguez}(2018)}]{var-rod:2018}
{Varniere}, P. \& {Rodriguez}, J. 2018, \apj, 865, 113

\bibitem[{{Wagoner} {et~al.}(2001){Wagoner}, {Silbergleit}, \&
  {Ortega-Rodr{\'{\i}}guez}}]{wag-etal:2001}
{Wagoner}, R.~V., {Silbergleit}, A.~S., \& {Ortega-Rodr{\'{\i}}guez}, M. 2001,
  AJ, 559, L25

\bibitem[{{Zhang} {et~al.}(2017){Zhang}, {Zhang}, {Yan}, {Fan}, \&
  {Liu}}]{zha-etal:2017}
{Zhang}, P., {Zhang}, P.-f., {Yan}, J.-z., {Fan}, Y.-z., \& {Liu}, Q.-z. 2017,
  APJ, 849, 9

\bibitem[{{Zhang} {et~al.}(2018){Zhang}, {Zhang}, {Liao}, {Yan}, {Fan}, \&
  {Liu}}]{zha-etal:2018}
{Zhang}, P.-f., {Zhang}, P., {Liao}, N.-h., {et~al.} 2018, APJ, 853, 193

\bibitem[{{Zhou} {et~al.}(2015){Zhou}, {Yuan}, {Pan}, \& {Liu}}]{zho-etal:2015}
{Zhou}, X.-L., {Yuan}, W., {Pan}, H.-W., \& {Liu}, Z. 2015, APJL, 798, L5

\end{thebibliography}

\end{document}